# Controlling the gain contribution of background emitters in few-quantum-dot microlasers

F. Gericke[1] , M. Segnon[2] , M. von Helversen[1] , C. Hopfmann[1] , T. Heindel[1] , C. Schneider[3], S. Höfling[3,4], M. Kamp[3] , A. Musiał[1,5] , X. Porte[1] , C. Gies[2] , and S. Reitzenstein[1]

[1] *Institut für Festkörperphysik, Technische Universität Berlin, 10623 Berlin, Germany*
[2] *Institute for Theoretical Physics, University of Bremen, 28334 Bremen, Germany*
[3] *Lehrstuhl für Technische Physik, Universität Würzburg, 97074 Würzburg, Germany*
[4] *School of Physics and Astronomy, University of St Andrews, St Andrews, KY16 9SS, UK*
[5] *Present address: Laboratory for Optical Spectroscopy of Nanostructures, Wrocław University of Science and Technology, Wrocław, Poland*

Corresponding author: S. Reitzenstein, phone: +49 30 314 79704, fax: +49 30 314 22064, Email: stephan.reitzenstein@physik.tu-berlin.de

**ABSTRACT**

We provide experimental and theoretical insight into single-emitter lasing effects in a quantum dot (QD)-microlaser under controlled variation of background gain provided by off-resonant discrete gain centers. For that purpose, we apply an advanced two-color excitation concept where the background gain contribution of off-resonant QDs can be continuously tuned by precisely balancing the relative excitation power of two lasers emitting at different wavelengths. In this way, by selectively exciting a single resonant QD and off-resonant QDs, we identify distinct single-QD signatures in the lasing characteristics and distinguish between gain contributions of a single resonant emitter and a countable number of off-resonant background emitters to the optical output of the microlaser.

We address the important question whether single-QD lasing is feasible in experimentally accessible systems and show that, for the investigated microlaser, the single-QD gain needs to be supported by the background gain contribution of off-resonant QDs to reach the transition to lasing. Interestingly, while a single QD cannot drive the investigated micropillar into lasing, its relative contribution to the emission can be as high as 70% and it dominates the statistics of emitted photons in the intermediate excitation regime below threshold.

**KEYWORDS** nanolasers, cavity QED, high-$\beta$ lasing, QD-micropillar cavities, single QD-laser, semiconductor lasers.

# INTRODUCTION

On the way towards the ultimate thresholdless semiconductor nanolaser[Noda2006], with only a single quantum dot (QD) as gain medium, the three main challenges are i) developing the required technology to realize such devices, ii) identifying lasing threshold, and iii) proving experimentally that a single QD is solely responsible for reaching it. Recent advances in material quality and the fabrication of semiconductor micro- and nanolasers allow one to enter the regime where only a few QDs, or even a single QD, provide the necessary optical gain[Xie2007],[Reitzenstein2008a], [Nomura2009], [Nomura2010],[Strauf2011], [Arakawa2012], [Liu2013]. So far, self-assembled QDs in semiconductor microcavities feature the highest optical quality in terms of oscillator strength, quantum efficiency and coherence properties[Birkedal2001], giving a chance to provide enough gain for lasing in the single-emitter limit. However, integrating a single self-assembled QD into a high-quality microcavity is a complicated task that requires sophisticated techniques, such as site-controlled growth[Pelucchi2007],[Surrente2009],[Pfau2009],[Schneider2009],[Strittmatter2012] or in-situ lithography[Lee2006],[Dousse2008],[Gschrey2013], which have been applied in the past to realize high-quality single-photon sources[He2016], [Sapienza2015], but up until now have not provided sufficient optical gain to reach the laser threshold.

In contrast, state-of-the-art QD-based microlasers have solely been based on self-assembled QDs placed randomly on the active area of the microlaser[Xie2007], [Reitzenstein2008a], [Nomura2010]. Most of these QDs can contribute to the output of the microlaser in an uncontrolled way, and only a small fraction of them have suitable spectral positions so they can be tuned through the cavity mode by, e.g., temperature tuning. Eventually, scenarios with only a single QD in spectral resonance (but not necessarily spatially matched) with the cavity mode are possible. Nevertheless, the requirements for such a single QD device to lase are very demanding. Even for a spontaneous emission factor (β-factor) close to unity, in which case spontaneous emission of the resonant emitter is almost solely directed into the laser mode, the light-matter coupling rate has to overcome the cavity loss rate at least by a factor of two[Gies2011]. In practice, it requires to combine cavities with a high quality factor (Q) and strong light-matter interaction, leading towards the coherent strong coupling regime[Reithmaier2004]. In this case the required high Q-factor microresonators with small mode volumes foster the illumination of the cavity mode by off-resonant QDs[Hennessy2007],[Press2007] which in turn has significant impact on the transition to lasing. Here, even spectrally far off-resonant emitters can couple to the cavity mode via different mechanisms, i.e., due to the interaction of QD excitations with acoustic phonons[RoyHughes2011], Auger-like scattering processes[Winger2009],[Florian2014] and Coulomb interaction with multi-exciton states[Laucht2010]. By this mechanisms, off-resonant QDs can feed the cavity mode within an energy range of ∼ 10 meV and contribute to lasing. Thus, a better understanding of the influence of individual in- and off-resonant QDs on the lasing behavior is needed and will be crucial for the design and operation of future nanolasers. This information is also an important contribution to ongoing very active discussion in the semiconductor community about the possibility for a single QD to provide enough gain to initiate and sustain lasing[Blood2013],[Coldren2013],[Ning2013],[Chow2014],[Gies2016],[Kreinberg2017]. Interestingly, and in spite of their central role, the influence of off-resonantly coupled QDs on the lasing behavior has not been described in a controlled and comprehensive way so far. We address this open issue, by using an advanced two-color excitation scheme with support from a microscopic laser theory. Our research gives important insight on the impact of background gain provided by off-resonant QDs in a regime, where the emission is dominated by a single resonant QD.

The structure under study is a high-quality low-mode volume GaAs-based QD-micropillar cavity containing a single layer of self-assembled QDs with an inhomogeneously broadened energy distribution of ≈ 50 meV. Our goal is to control the gain contribution of off-resonantly coupled QDs in our microlaser and to distinguish their influence and the lasing behavior from that of the desired resonant QD. This allows us to identify fingerprints of different gain contributions to the laser output and, as a result, distinguish between devices with only one QD and with a few QDs constituting the gain of the microlaser, simply by varying the relative intensity of two excitation lasers. We do so by using a two-color excitation scheme: The target QD gain is selectively addressed by resonant excitation of its spectrally narrow p-shell resonance, while the gain of the off-resonantly coupled QDs is controlled simultaneously by above band excitation. Thereby, the ratio between the two different excitation powers is used to control the relative contribution of the off-resonant emitters to the device output.

In general, nanolasers operating in the high-β regime do not show a pronounced and typical laser characteristics in the input-output curve[Chow2014]. Therefore, the identification of the lasing threshold for a nanolaser is a challenging task that usually requires to take into account different emission characteristics[Choi2007], [Yokoyama1992], [Assmann2009], [Nomura2007], [Hostein2010], [Hachair2011], [Strauf2011], [Lu2012]. In this context, we apply a microscopic semiconductor laser model to precisely determine the threshold of the investigated microlaser in the different experimental scenarios. Following this approach, we obtain a comprehensive understanding of the laser's threshold and its β-factor, which in our experiment is a function of the background gain contribution due to the different coupling coefficients of the resonant QD and the non-resonant background emitters.

**MATERIALS AND METHODS**

**SAMPLE PROPERTIES:** For our present study it is crucial that the QD in resonance couples efficiently to the cavity mode and that the contribution of the off-resonant emitters to the laser output is non-negligible. Therefore, we have used a high quality factor ($Q \approx 15000$) low-mode volume micropillar with a diameter of 1.8 µm, maximizing the light-matter coupling strength between the exciton transition of the resonant QD and the fundamental cavity mode. The gain medium consists of a single layer of self-assembled InGaAs QDs, with an Indium content of about 40 % and an areal density of $10^{10} cm^{-2}$ in the center of a GaAs λ-cavity. These QDs feature a large oscillator strength, which in combination with the low mode-volume micropillar ensures pronounced light-matter interaction that facilitates reaching the strong coupling regime[Press2007] with pronounced single QD lasing effects[Gies2016]. On top (bottom) of the central GaAs cavity 26 (30) pairs of AlAs/GaAs layers acting as high reflective distributed Bragg reflectors (DBR) were grown. The micropillar was realized by high-resolution electron-beam lithography and plasma etching. A scanning electron micrograph (SEM) of a processed free standing micropillar is shown in Fig. 1(a). The λ-cavity is visible in this picture as the thicker central horizontal section. The sample was cleaved to gain optical access to the micropillar cavity from the side (in the direction perpendicular to the micropillar axis). This enables direct and wavelength-independent excitation of the QDs [Ates2009]. For further details on the sample layout and processing we refer to Ref. [Reitzenstein2007].

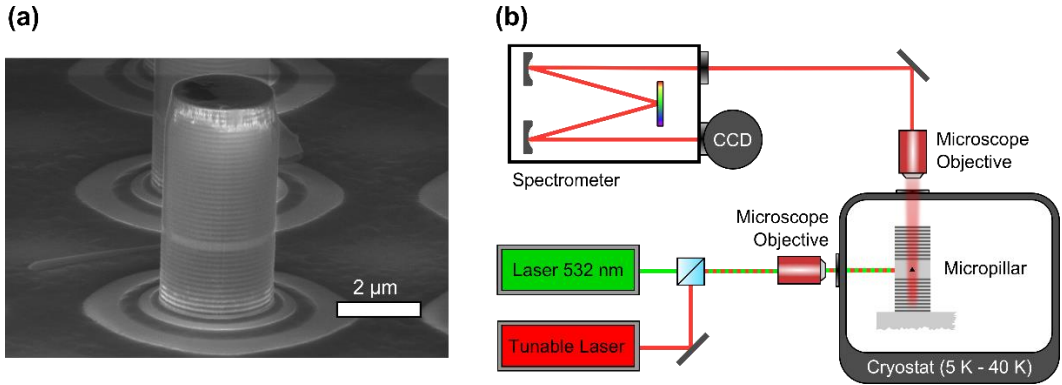

**Figure 1.** (a) Scanning electron micrograph (SEM) of an exemplary processed free standing micropillar. The bottom distributed Bragg reflector (DBR) is only partly etched. (b) Sketch of the experimental micro-photoluminescence (µPL) setup with a configuration of lateral excitation and axial detection.

To gain insight into the lasing characteristics of the QD-micropillar structure, its optical output was studied as a function of excitation power using the micro-photoluminescence (µPL) setup schematically shown in Fig. 1(b). This setup has a perpendicular configuration of the excitation and the detection paths. The main advantage of side-excitation here is that the laser light is not (partially) blocked by the stop-band of the top DBR[Musial2015]. Therefore, an efficient and homogeneous, i.e. wavelength-independent, excitation of the QDs can be realized. Furthermore, the perpendicular excitation and detection paths provide a natural rejection of a big fraction of the pump lasers' light in detection – an advantage that is particularly important for pumping wavelengths close to the micropillar's resonance frequency. To selectively excite a single QD resonant with the cavity mode, we apply a quasi-resonant p-shell excitation scheme using a tunable semiconductor infrared (IR) laser with linewidth below 100 kHz (0.41 neV). The optical above-band excitation of the sample is carried out by a frequency doubled Nd:YAG laser emitting at a wavelength of 532 nm (further referred to as green excitation). The output power of each laser can be independently attenuated via a set of variable density filters before they are combined on a beam-splitter and focused on the sample by a lateral objective featuring high numerical aperture of 0.4 and long working distance of 20 mm. The sample is mounted in a variable temperature He-flow cryostat and kept at constant temperature of 25 K for most of the experiments. The far-field emission of the fundamental cavity mode is in perpendicular direction to the excitation path.

Based on the areal QD density of the wafer, we estimate an amount of $\simeq$ 250 dots within the active layer of a micropillar with 1.8 µm diameter. Due to the self-assembled character of QD growth, there is a high variability in the QD emission energy and the spatial position. Nevertheless, about 5 QD lines are in the spectral proximity of the lasing mode and can be studied by fine-tuning with respect to the cavity mode. In the present case, the chosen QD excitonic transition couples strongly to the cavity mode at a resonance temperature of 25 K. At the same time, the spectral density is high enough to provide enough background gain to overcome the lasing threshold.

Above-band excitation is used to excite the background emitters. Carriers are generated in the barrier material, from where they are captured equally into all QDs irrespective of their transition energies. In contrast, to address a target QD selectively either resonant (s-shell) or quasi-resonant (p-shell) excitation scheme has to be employed. We choose p-shell excitation for most of the experiments because, in comparison with s-shell excitation, laser stray-light suppression is less demanding. To determine the energy of the p-shell for QDs in the micropillar of interest, we perform an excitation wavelength-dependent measurement, i.e. micro-photoluminescence excitation (µPLE), at low excitation powers (not shown here). Whenever the laser energy is resonant with a p-shell (or another higher lying resonance) of a QD, we see a sharp maximum in the emission intensity at the energy of this QD and the cavity mode due to efficient pumping of the corresponding QD followed by the excitation transfer into the cavity mode due to off-resonant QD-cavity coupling. The response of the mode gets stronger the less detuned a QD is with respect to the cavity due to more efficient non-resonant cavity feeding. We selected the QD with the strongest p-shell resonance. It can be tuned into resonance with the laser mode and exhibits a splitting between the s-shell and the p-shell of ≈ 13 meV. This splitting is small in comparison to typical values of ∼ 25-30 meV for standard In(Ga)As QDs [Cusack1996], [Brounkov1998], [Narvaez2005], which is in agreement with an enhanced in-plane spatial extension of investigated QDs.

**OPTICAL CHARACTERIZATION:** First, we evaluate the influence of the background emitters on the microlaser characteristics by examining the power-dependent emission spectra in two limiting cases: selective p-shell excitation of a target QD in resonance with the cavity mode (Fig. 2(a)) and non-selective above- band excitation with a green laser of all QDs in the gain medium (Fig. 2(b)).

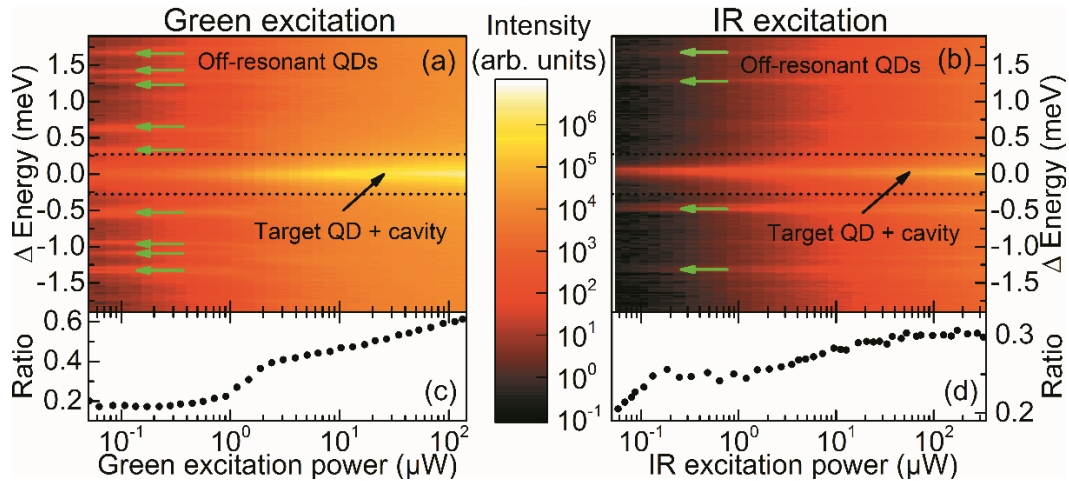

**Figure 2.** Power-dependent emission spectra for the case of only above-band (green) excitation (a) and only p-shell (IR) excitation of the target QD in resonance with the cavity mode (b). The energy difference is relative to the central energy of the cavity mode emission at high excitation powers. The off-resonant QDs are marked by green arrows. Panels (c) and (d) show the ratio of the integrated total single QD and cavity mode intensity (area between the black dotted lines in (a) and (b)) and the residual area of the spectrum for the two respective excitation schemes.

The qualitative differences between the two cases are visible in the two panels of Fig. 2. Using above-band excitation (Fig. 2(a)), the QD emission lines (indicated by green arrows) exhibit larger linewidths at low excitation and broaden strongly with increasing excitation power. At high excitation ≥ 5 µW, the spectrum is dominated by the cavity mode and strong broadband background so that single QD emission lines cannot be resolved anymore. These observations can be attributed to the fact that a large number of high-energy carriers are created in the whole structure that undergo multi-stage relaxation processes into the lowest energy states in the QDs. At higher excitation powers, when the confined states in the QDs are saturated, the recombination takes place from higher-energy states in the structure (wetting layer or GaAs barrier material). This constitutes an additional background that contributes to the output of the micropillar, and it eventually gets stronger than the emission from single QDs experiencing saturation. In contrast, Fig. 2(b) depicts the spectral dependence when only the p-shell of the selected QD is pumped. Due to a lower amount of carriers and less decoherence in the system, QD lines are narrower and do not broaden significantly with increasing excitation power, so that they can be individually resolved in the whole covered excitation range. Interestingly, even though we are using quasi-resonant excitation of a target QD, off-resonant QDs are still visible in the spectrum. This observation can be explained by the strong light-matter coupling in the structure leading to cavity mediated coupling between the QDs as discussed above. In this process, a target QD emits a photon that is stored in the cavity and afterwards transferred via one of the off-resonant coupling mechanisms, i.e. interaction with acoustic phonons, Auger-like scattering or Coulomb interaction with multi-excitonic states, to an off-resonant QD. An analogue effect was previously observed in resonance fluorescence experiments on the same sample[Hopfmann2016]. Another possibility is that due to relatively shallow confining potential of the investigated QDs, the p-shell overlaps energetically with the tail of the density of states in the wetting layer (WL). This would result in non-zero probability of scattering carriers created in the p-shell state out of the QD towards the WL[Musial2014] (instead of relaxing to the s-shell of the target QD). Since carriers in the WL can be captured into any of the QDs in the active region, this effect would be a detrimental factor to the selectiveness of our quasi-resonant excitation scheme.

To further quantify the difference in the response of the system under the two applied excitation schemes, we evaluate the single QD and the background emitters' contribution to the spectra in terms of integrated intensities. For this purpose, we calculate the ratio between the QD in resonance with the cavity mode (selected range is marked as dotted lines in Figs. 2(a) and (b)) and the integrated intensity of the rest of the presented spectrum (outside the dotted lines). Figure 2(c) depicts the system response under non-resonant excitation. The ratio shows a strong nonlinear increase in favor of the cavity mode contribution starting from $P \approx 1$ µW showing that most of the emission is funneled into the cavity mode and contributes to the microlaser output. This can be attributed to reaching the onset of stimulated emission and resembles a typical input-output laser characteristics. With increasing excitation power, the cavity is more effectively fed by the off-resonant emitters, which is reflected in a decreasing contribution of their intensity to the total intensity – a behavior that we consider as fingerprint of lasing action. Figure 2(d) shows the described ratio for the quasi-resonant IR pumping scenario. Noteworthy, at low excitation powers, under p-shell excitation the cavity is fed more efficiently than when the above-band pump is applied, as it is indicated by the $\approx 7$ % higher value of the ratio at low pump powers. This behavior can be attributed to strong coupling of the single QD in resonance to the cavity mode. The steeper initial increase in the intensities ratio is a fingerprint of the single-QD nonlinearity proving that indeed in this excitation range the contribution of the background emitters is negligible. The subsequent power-dependent evolution differs strongly from the above-band excitation scenario depicted in Fig. 2(c). For the p-shell excitation of the target QD, the ratio stays almost constant within ≈ 5-10 % variation and does not scale proportionally to the excitation power. This supports the interpretation that excitation of the system comes almost exclusively from a single emitter (at low excitation powers), which undergoes saturation for intermediate to high excitation powers.

**MICROSCOPIC LASER MODEL FOR RESONANT QD AND BACKGROUND EMITTERS:** To gain further insight in the presented input-output curves and their interrelation with single-QD lasing, we employ a theoretical laser model that accounts for the semiconductor gain medium. The gain consists of a single QD resonant with the cavity mode (referring quantities labeled $\xi$ = QD) and $N_{BG}$ background emitters ($\xi$ =BG). Our microscopic model is based on the approach introduced in [Gies2007] and consists of a set of coupled dynamical equations derived from the Hamiltonian for the electronic states of the QD emitters, photons of the quantized electromagnetic field, and the interaction between QD excitations and photons in the laser mode. A set of coupled dynamical equations is derived for the intracavity mean photon number ($\langle b^\dagger b \rangle$), and carrier populations in the resonant QD ($f_{s,QD}^{e,h}$) and the off-resonant QDs ($f_{s,BG}^{e,h}$):

$$\left(\hbar \frac{d}{dt} + 2\kappa\right)\langle b^\dagger b\rangle = 2|g_{QD}|^2 \langle b^\dagger v^\dagger c\rangle_{QD} + 2N_{BG}|g_{BG}|^2 \langle b^\dagger v^\dagger c\rangle_{BG} \quad (1)$$

$$\hbar \frac{d}{dt} f_{s,QD}^{e,h} = -2|g_{QD}|^2 \langle b^\dagger v^\dagger c\rangle_{QD} + \mathcal{R}_{nl}(\beta_{QD}) + \mathcal{R}_{p\to s}^{e,h}(P_g, P_{IR}) \quad (2)$$

$$\hbar \frac{d}{dt} f_{s,BG}^{e,h} = -2|g_{BG}|^2 \langle b^\dagger v^\dagger c\rangle_{BG} + \mathcal{R}_{nl}(\beta_{BG}) + \mathcal{R}_{p\to s}^{e,h}(P_g, P_{IR}) \quad (3)$$

Here $\kappa$ is the cavity loss rate, $g_\xi$ denotes the coupling constant of the QD in resonance or that of the off-resonant emitters, and the operators $c^\dagger, v^\dagger$ annihilate (create) a carrier in the s-shell

conduction- or valence-band state of each emitter. Operators $b^\dagger$ address photons in the laser mode. The rate $\mathcal{R}^{e,h}_{p\to s}$ describes the creation of excited carriers in the laser levels via scattering that follows excitation from the two pump sources, green and infrared, with respective pump powers $P_{IR}$ and $P_g$. These excited carriers are created into the energetically higher p-states via a relaxation-time approximation. The spontaneous recombination of carriers into nonlasing modes is given by the rate $\mathcal{R}_{nl}$ that depend on the β-factors of the resonant QD ($\beta_{QD}$) and the background emitters ($\beta_{BG}$). The dynamics of Eqs. (1)-(3) is determined by a balance of these interaction processes with the environment and the light-matter interaction of the single resonant and $N_{BG}$ background QDs via photon-assisted polarizations

$$\left(\hbar \frac{d}{dt} + \kappa + \Gamma_\xi\right)\langle b^\dagger v^\dagger c\rangle_\xi = f^e_{s,\xi} f^h_{s,\xi} + (1 - f^e_{s,\xi} - f^h_{s,\xi})\langle b^\dagger b\rangle \qquad (4)$$

with the dephasing $\Gamma_\xi$ associated with the QD transitions resonant with the laser mode. This equation contains the spontaneous-emission contribution $\propto f^e f^h$ and the stimulated emission or absorption term proportional to the intra-cavity mean photon number that also appears in rate–equation theories. While the rate equations could be obtained by adiabatically eliminating the photon-assisted polarizations, we calculate the full dynamics and the dynamics of higher-order carrier-photon correlations $\delta\langle b^\dagger b\, c^\dagger c\rangle$, $\delta\langle b^\dagger b v^\dagger v\rangle$, $\delta\langle b^\dagger b^\dagger b v^\dagger c\rangle$, and $\delta\langle b^\dagger b^\dagger b b\rangle$ as described in the Supporting Information. These equations allows us to calculate the second-order photon-correlation function at zero time delay $g^{(2)}(0)$ which contains information on the statistical properties of the emission differentiating between single-photon character ($g^{(2)}(0) < 1$), thermal ($g^{(2)}(0) = 2$), and coherent ($g^{(2)}(0) = 1$) emission.

We determine the light-matter coupling-strength $g_\xi$ and the β-factor individually for the resonant and off-resonant case on the basis of experimental data obtained under purely green or IR excitation as shown in Fig. 3. See supplementary information for further details. These parameters are used in all following calculations and only the pump rates are varied to obtain the two-color excitation plots.

To further understand the nature of excitation in our system, it is important to note that the two components of the gain in our laser model (resonant QD and background emitters) are coupled via the common light field of the cavity. This leads to the effect that the resonant QD can in fact be indirectly excited by background excitation by reabsorbing cavity photons that were emitted from the detuned background emitters, and vice-versa. It is therefore not possible to separate the system into resonant and background parts other than by switching off the corresponding light-matter coupling completely, a possibility that is reserved to theory alone and is illustrated in Fig. 6(a).

Emitters that are spectrally and spatially detuned from the cavity mode naturally possess a weaker light-matter coupling strength and, thus, a lower β-factor than the single QD in resonance with the cavity mode. Consequently, the β-factor of a system consisting of resonant and background emitters depends sensitively on the contribution of each. It is possible to quantify the β-factor from Eqs. (2)-(4) by considering only the spontaneous-emission contributions and solving Eq. (4) adiabatically. In this case, an effective β-factor can be expressed as (see Supporting Information)

$$\beta_{eff} = \frac{\beta_{QD}}{1+\frac{N\langle b^\dagger v^\dagger c\rangle_{BG}}{\lambda\langle b^\dagger v^\dagger c\rangle_{QD}}} + \frac{\beta_{BG}}{1+\frac{\lambda\langle b^\dagger v^\dagger c\rangle_{QD}}{N\langle b^\dagger v^\dagger c\rangle_{BG}}}, \quad \lambda = \frac{\beta_{BG}}{\beta_{QD}}\left|\frac{g_{QD}}{g_{BG}}\right|^2 \tag{5}$$

In the limit of vanishing contributions from background emitters, $\beta_{eff}$ takes on the high β-value of the resonant emitter, whereas a significantly lower $\beta_{eff} = \beta_{BG}$ is observed in the case of a dominating background. Via λ, not only the number of background emitters enters, but also the respective coupling strength, taking into account the weaker coupling of detuned emitters.

**RESULTS AND DISCUSSION**

**BACKGROUND CONTROLLED LASING IN A QD-MICROPILLAR SYSTEM:** The actual experimental input-output curves for the two limiting cases, i.e., above-band excitation and p-shell excitation of the target QD, respectively, are presented with symbols in Fig. 3(a) and (b). In the first case, in which all QDs are excited and can contribute to the gain, the input-output dependence shows the pronounced s-shape that is characteristic for the onset of stimulated emission in microlasers. In contrast, the p-shell excitation scenario results in nearly linear behavior over the whole measured range. Noteworthy, saturation at some point on the input-output curve would be expected for this latter scenario, but is not observed. Further experiments (cf. Supplementary information) in a resonant pumping scenario demonstrate that the QD in resonance is indeed saturating under strong p-shell excitation. However, the fact that we do not observe saturation in the input-output curve (cf. Fig. 3(b)) shows that indeed the off-resonant emitters are also (unintentionally) excited and can even dominate the output of the QD-micropillar at high pump rates.

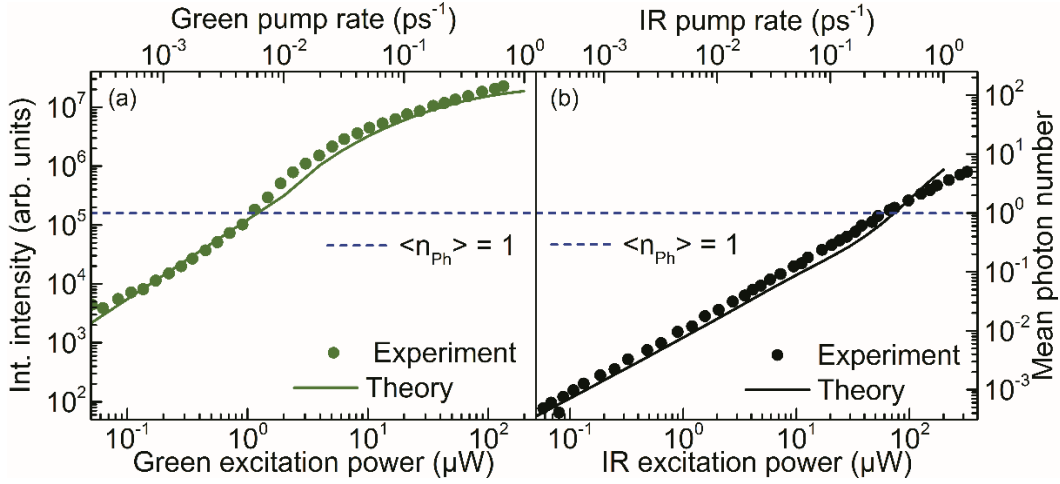

**Figure 3.** Experimental (dots) and theoretical (line) input-output characteristics for (a) only exciting above-band (green) and (b) only exciting the p-shell of the target QD (IR). The experimental points of the target QD in resonance with the cavity mode were calculated by integrating the Rabi-doublet area of the spectra, delimited with dotted lines in Figs. 2(a) and (b). In both panels the laser threshold, defined in the numerical model as $\langle n_{Ph} \rangle = 1$, is indicated with a dashed blue line.

The clearly different behavior between both panels in Fig. 3 demonstrates that our two-color excitation scheme can be used to understand and tailor the output characteristics of a few-QD semiconductor microlaser, including the β-factor, by selective manipulation of the resonant and background gain contribution.

Up to now, the two limiting cases of exciting dominantly the single target QD or all QDs in the micropillar, have been presented. Now, we analyze the transition between them by gradually unbalancing between the two different pumps and continuous measurement of the QD-micropillar output characteristics. The results of full two-color excitation measurements are shown as excitation maps in Fig. 4 as obtained from experiment (panel (a)) and from the theoretical model (panel (b)).

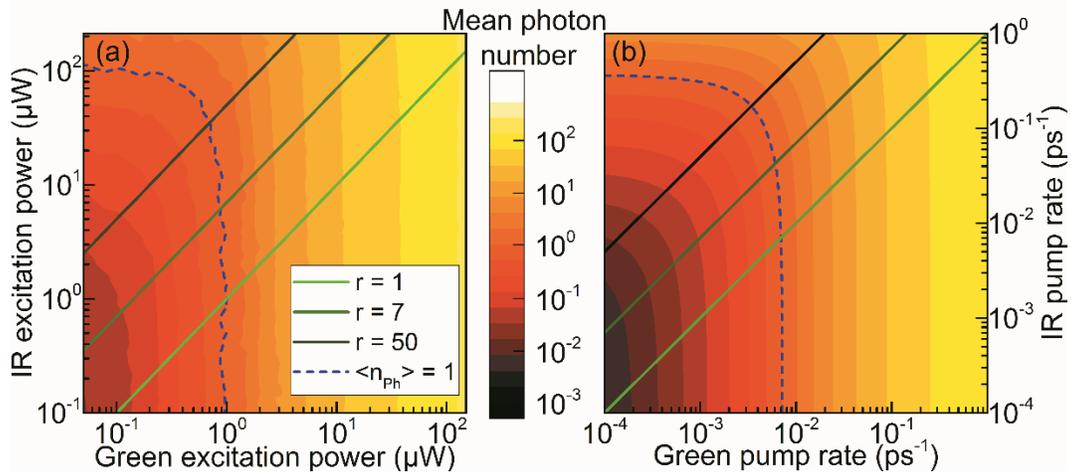

**Figure 4.** Experimental (left) and modelled (right) dependence of the emission intensity on the excitation conditions in the two-color excitation scheme. The blue dashed line in both panels represents the lasing threshold according to the usual definition $\langle n_{Ph} \rangle = 1$ for microlasers. Above this line (lighter colors areas) the micropillar output is dominated by stimulated emission.

The horizontal axis represents the strength of the above-band excitation. Increasing the corresponding pump-rate corresponds mainly to increased excitation of the off-resonant emitters in the micropillar. In the vertical direction, p-shell excitation of the target single QD is increased. The blue dashed line in the left panel corresponds to the usual definition of threshold power ($\langle n_{Ph} \rangle = 1$) determined from comparison with results of the numerical modeling. Noteworthy, the qualitative agreement between the experimental and the theory maps is very high. The presented maps prove that the difference between input-output curves for the limiting cases is not related to different scaling factors for the excitation power but indeed to the fact that achieving lasing conditions with a single QD gain is rather challenging.

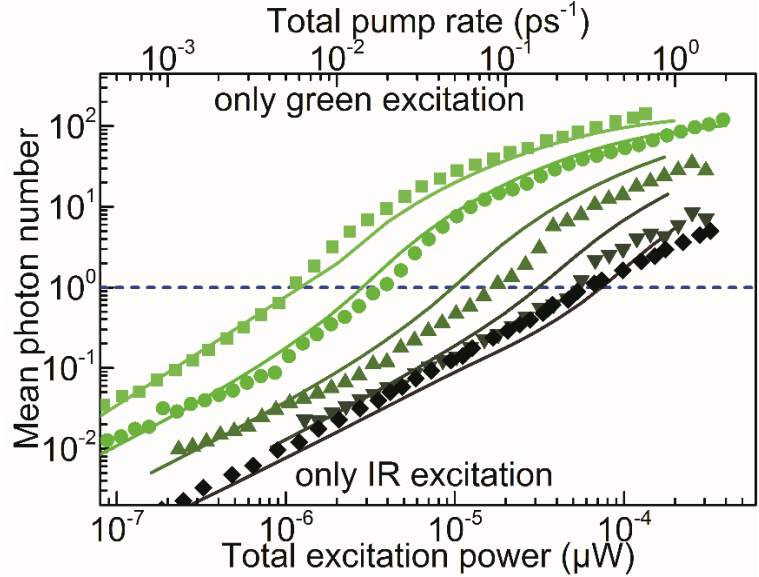

**Figure 5.** Diagonal cross-sections through the 2D map (indicated with the corresponding colors in the previous figure) for three different green to IR excitation power ratios r = 1, 7 and 50 (corresponding to constant relative contribution of the off-resonant emitters to the gain) together with the previously shown limiting cases are plotted versus the sum of both excitation powers. The symbols correspond to the experimental data of the 2D map and the lines to the calculations shown in Fig. 4. The ratio r is defined as the excitation power of the green laser divided by the excitation power of the IR-laser. The blue dashed line indicates the lasing threshold ($\langle n_{Ph} \rangle = 1$) as extracted from the theory fit.

To visualize the change in the shape of the input-output curve, diagonal cross-sections through the 2D map are presented in Fig. 5 at positions indicated by the colored solid lines in Fig. 4(a). The upper- and lower-most input-output curves correspond to the two limiting cases shown in Figs. 3 (a) and (b). The diagonal cross-sections correspond to the input-output characteristics at fixed ratios r = 1, 7, 50 between both excitation powers, i.e., constant contribution-percentage of the off-resonant emitters. The input-output curves in Fig. 5 are plotted against the sum of both excitation powers. The complementary theory curves are also plotted in the same panel together with an horizontal (blue dashed) line associated with a mean photon number $\langle n_{Ph} \rangle = 1$, indicative for the lasing threshold. It can be clearly seen that the increase of the off-resonant emitter-contribution causes the s-shape in the transition regime to become more pronounced and the threshold position shifts towards lower total excitation powers. Interestingly, the higher fraction of light coupled into the cavity mode from the background emitters with less-ideal light-matter coupling strength simultaneously degrades the effective β-factor of the emission. We quantify this effect on the basis of the effective β-factor defined in Eq. (5), which is evaluated numerically. The result is shown in Fig. 6(a): The maximal achievable effective β-factor of 0.37 in case of dominant p-resonant excitation is still more than two times smaller than the β-factor for the target resonant QD, which we extract to be 0.9 from matching the result shown in Fig. 3 for selective IR excitation of the single QD only (without any background emission). This indicates that even a weak above band excitation with a pump rate as low as $10^{-4}$ ps$^{-1}$ introduces significant background illumination of the cavity mode. Noteworthy, for only above-band excitation of the system and in the strong excitation regime, the β-factor is drops to values close to $\beta_{BG}= 0.25$, evidencing the dominant role of the background emitters in this range. Only in the regime of intermediate IR pump rates, the single QD gains a meaningful contribution so that its fingerprints become visible in the microlaser characteristics. In this low excitation regime, these characteristics distinguish between a microlaser with only a single QD gain and a multi-QD laser.

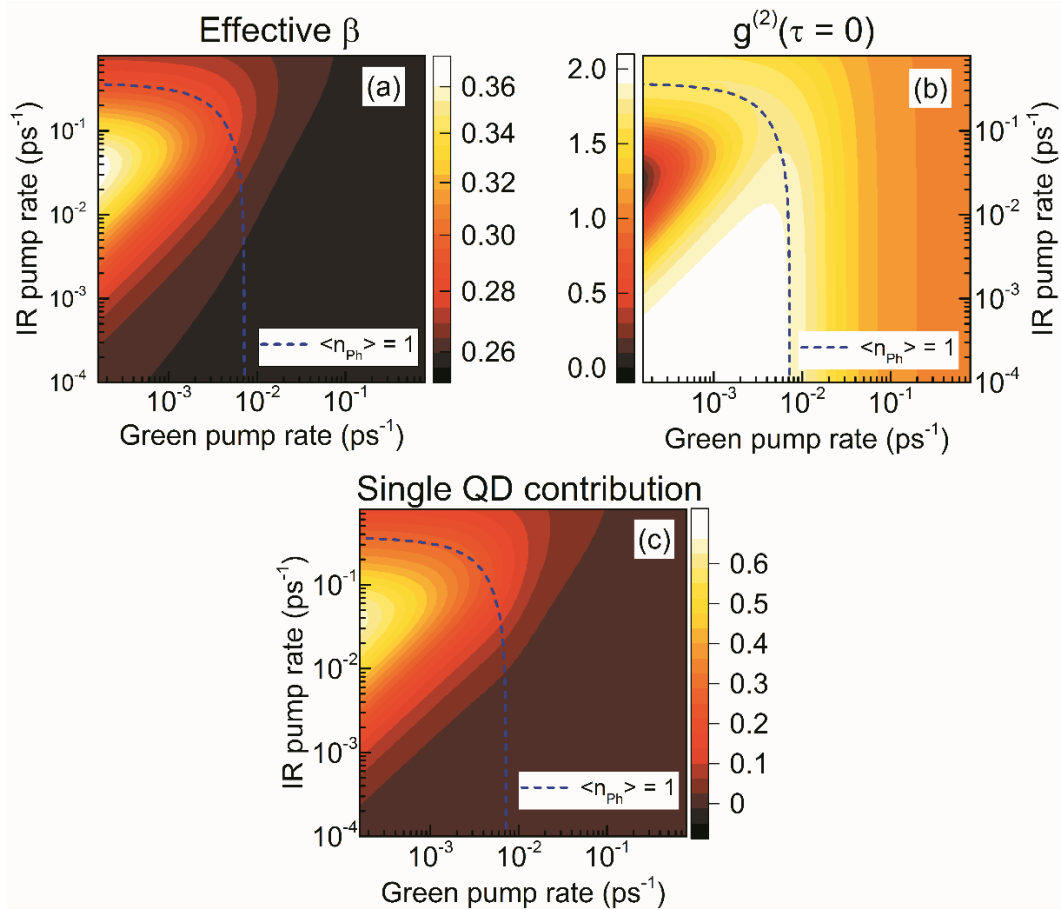

**Figure 6.** With the two-color pump rates on both axes, numerical results are shown for (a) the effective β-factor, (b) $g^{(2)}(\tau = 0)$, and (c) the contribution of the single target QD to the total micropillar output, which is evaluated as the relative difference between the total emission intensity of the full system, and the total emission of the background emitters for parameters corresponding to the system investigated experimentally. In all graphs the laser threshold ($\langle n_{Ph} \rangle = 1$) is marked by a blue dashed line.

Similar regions can be identified in the photon statistics. The calculated $g^{(2)}(\tau = 0)$ map is presented in Fig. 6(b). Also in this case the behavior is non-monotonic with $P_{IR}$: Generally, in the low excitation regime, $g^{(2)}(\tau = 0)$ of 2 reflects the thermal character of the emission from the QD ensemble. This at first sight unexpected behavior is explained by the fact that a small fraction of $P_{IR}$ also drives the background. In a realistic case of exciting 100 background emitters by 1%, their spontaneous emission becomes comparable to the contribution of the single resonant QD. Then light is thermal, because the SQD becomes "part of the ensemble". With increasing carrier population, i.e. high $P_{IR}$ spontaneous emission becomes faster, as it is proportional to the populations ($f_e$ x $f_h$), and once inversion is reached, stimulated emission sets in for the single QD but not for the background emitters. Therefore, the single QD contribution dominates only at higher excitation the statistical properties of the emission, revealing non-classical behavior and antibunching. Interestingly, even if there was not fraction of $P_{IR}$ driving the background, we would observe a similar effect, because photons emitted into the cavity by the single QD were re-absorbed by the background QDs, so that even then, the emission would be thermal at very low $P_{IR}$. At high incoherent excitation (using $P_G$) coherent emission is reached at pump rates of about 0.1/ps. Since a small fraction of the p-shell excitation also drives the background emitters, coherent emission can also be approached when $P_{IR}$ is further increased, even though the single QD alone does not provide sufficient gain to cross the laser threshold.

Both the effective β-factor and $g^{(2)}(\tau = 0)$ dependencies on the excitation power can be traced back to the relative contribution of the single QD and background emitters to the output of the microlaser presented in Fig. 6(c). This intensity map presents the relative contribution of a single QD to the emission evaluated as a relative difference between the total emission (background emitters and the single QD) and the background emitter's emission only (in which case the single QD is removed in the modelling). This cannot be realized in our experiments, as the presence of the single QD is noticeable even if is not excited directly. Thus, the theoretical analysis gives important insight beyond the experimentally accessible regime and is very informative as it separates the two contributions to the microlaser output and shows up to 70 % increase in emission intensity enhancement due to only a single emitter (for details see supplementary information). Additionally, regions where the emission of a single QD shows saturation (at high IR excitation pump rates in excess of 0.1 ps$^{-1}$). In their sum, the isolated contribution of the SQD and the effective β-factor provide important insight into the interplay of resonant and background contributions in a nanolaser that can operate close to the ideal regime of single-emitter lasing. This insight could not be obtained from $g^{(2)}(\tau = 0)$ alone, which is a more intricate quantity as it reflects the properties of the photons in the cavity, rather than their origin. At the same time, the autocorrelation function demonstrates that a single device can be operated in any regime of non-classical, coherent, or thermal emission by choosing the resonant (IR) and background (green) excitation to realize any point in the two-color maps.

**CONCLUSIONS**

We have presented a comprehensive experimental and theoretical analysis of the relative gain contribution of a single resonant emitter and non-resonant background emitters which are off-resonant in the single-QD lasing regime. Experimentally, this study is enabled by a two-color excitation scheme in a lateral excitation/axial detection experimental configuration in which QDs can be excited directly at any wavelength. The contribution of the off-resonant QDs is controlled optically by above-band excitation, meanwhile the single QD in resonance is excited selectively via its p-shell. This advanced excitation scheme allowed us to demonstrate a transition between a device with characteristics similar to those of a macroscopic laser with QD-ensemble gain, and a microlaser fed by a very limited and discrete gain which requires a quantum-optical description of interaction between a two-level system and photons in the cavity. Therefore, our study provides important insight into the operation of high-quality microlasers close to the limiting case of the thresholdless single emitter laser. In particular, it allows us to distinguish between a single- and a multiple-QD laser, a task which cannot be done solely based on the input-output characteristics.

We reveal that a dominant single QD role leads to a higher effective β-factor because of the dominating single-QD contribution to the emission. This is a key aspect of our work, which shows that, in contrast the usual understanding, the β-factor is not constant for a given microcavity system, but depends on and can be controlled by the specific excitation conditions determining the effective gain. The dominant single QD role is also evidenced in the intermediate regime in the photon statistics. Both, the lasing threshold and the effective β-factor, strongly depend on non-resonant gain contribution. Nevertheless, even if the efficiency of the spontaneous emission coupling to the lasing mode is degraded by off-resonant emitters, lasing conditions can be reached in our system due to additional emitters.

The developed experimental approach is a very powerful technique enabling continuous change of the output characteristics of a single microlaser device using selective excitation of its gain which constitutes an alternative to more complicated schemes, where precise or even deterministic control of the position, number and optical characteristics of QDs in the active material during growth or processing is utilized. Our analysis demonstrates that the off-resonant QDs lower the threshold power and result in restoring a pronounced s-shape in the input-output curve, but simultaneously cause a drop in the effective β-factor in QD-based micropillar lasers. Therefore, the contribution of the non-resonant QDs can be used to control and tailor those two correlated laser parameters. As such our work provides important insight into the relative contribution of a resonant emitter and non-resonant background emitters on the emission properties of a microlaser, which will be of high relevance for the further development of micro- and nanolasers towards the ultimate thresholdless single quantum dot laser.

**ACKNOWLEDGEMENTS**

The research leading to these results has received funding the European Research Council under the European Union's Seventh Framework ERC Grant Agreement No. 615613 as well as from the German Research Foundation via Grant-No.: Re2974/10-1, Gi1121/1-1. We gratefully thank M. Sommer for technical assistance.

**Figure captions**

**Figure 1.** (a) Scanning electron micrograph (SEM) of an exemplary processed free standing micropillar. The bottom distributed Bragg reflector (DBR) is only partly etched. (b) Sketch of the experimental micro-photoluminescence µPL) setup with a configuration of lateral excitation and axial detection.

**Figure 2.** Power-dependent emission spectra for the case of only above- band (green) excitation (a) and only p-shell (IR) excitation of the target QD in resonance with the cavity mode (b). The energy difference is relative to the central energy of the cavity mode emission at high excitation powers. The off-resonant QDs are marked by green arrows. Panels (c) and (d) show the ratio of the integrated total single QD and cavity mode intensity (area between the black dotted lines in (a) and (b)) and the residual area of the spectrum for the two respective excitation schemes.

**Figure 3.** Experimental (dots) and theoretical (line) input-output characteristics for (a) only exciting above-band (green) and (b) only exciting the p-shell of the target QD (IR). The experimental points of the target QD in resonance with the cavity mode were calculated by integrating the Rabi-doublet area of the spectra, delimited with dotted lines in Figs. 2(a) and (b). In both panels the laser threshold, defined in the numerical model as $\langle n_{Ph} \rangle = 1$, is indicated with a dashed blue line.

**Figure 4.** Experimental (left) and modelled (right) dependence of the emission intensity on the excitation conditions in the two-color excitation scheme. The blue dashed line in both panels represents the lasing threshold. Above this line (lighter colors areas) the micropillar output is dominated by stimulated emission.

**Figure 5.** Diagonal cross-sections through the 2D map (indicated with the corresponding colors in the previous figure) for three different green to IR excitation power ratios r = 1, 7 and 50 (corresponding to constant relative contribution of the off-resonant emitters to the gain) together with the previously shown limiting cases are plotted versus the sum of both excitation powers. The symbols correspond to the experimental data of the 2D map and the lines to the calculations shown

in Fig. 4. The ratio r is defined as the excitation power of the green laser divided by the excitation power of the IR-laser. The blue dashed line indicates the lasing threshold ($\langle n_{Ph} \rangle = 1$) as extracted from the theory fit.

**Figure 6.** With the two-color pump rates on both axes, numerical results are shown for (a) the effective β-factor, (b) $g^{(2)}(\tau = 0)$, and (c) the contribution of the single target QD to the total micropillar output, which is evaluated as the relative difference between the total emission intensity of the full system, and the total emission of the background emitters for parameters corresponding to the system investigated experimentally. In all graphs the laser threshold ($\langle n_{Ph} \rangle = 1$) is marked by a blue dashed line.

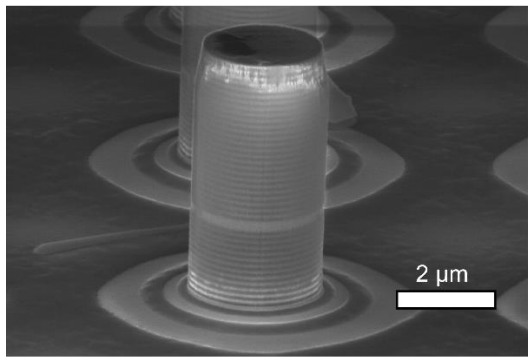 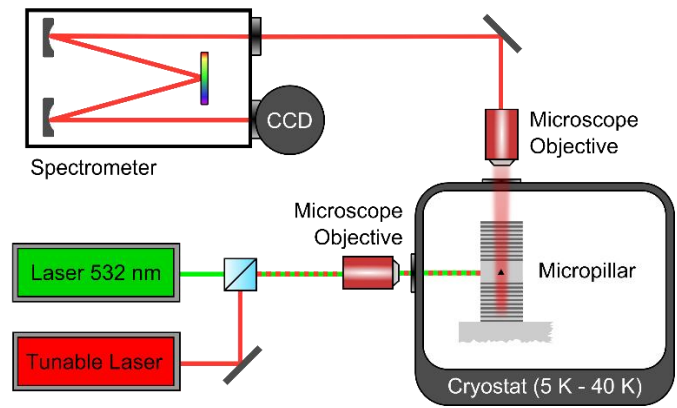

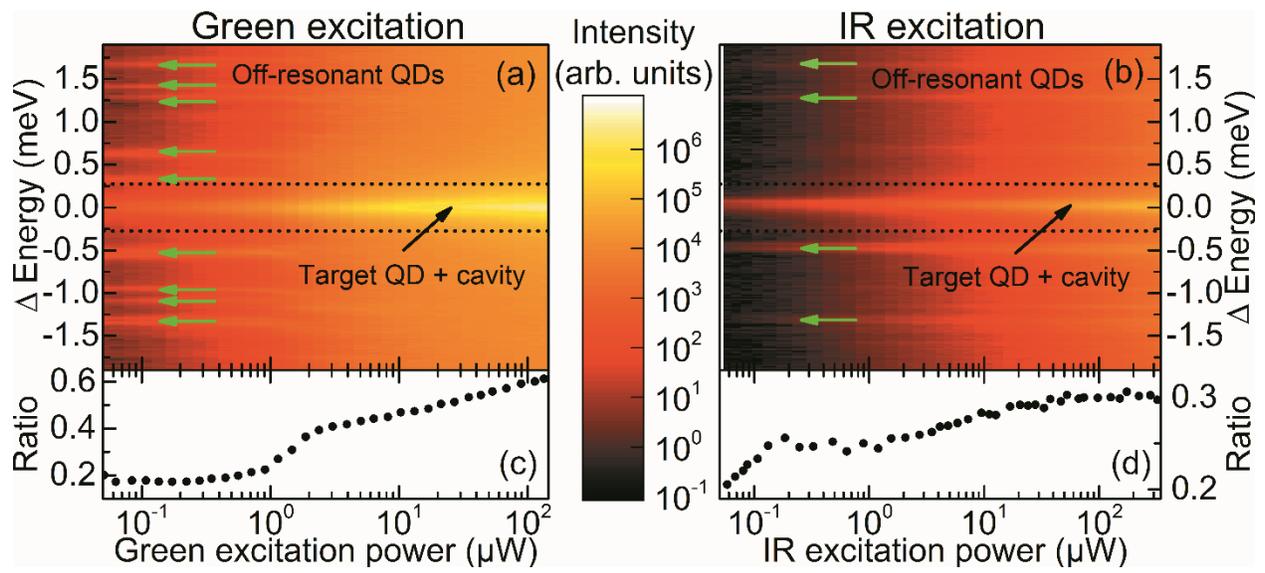

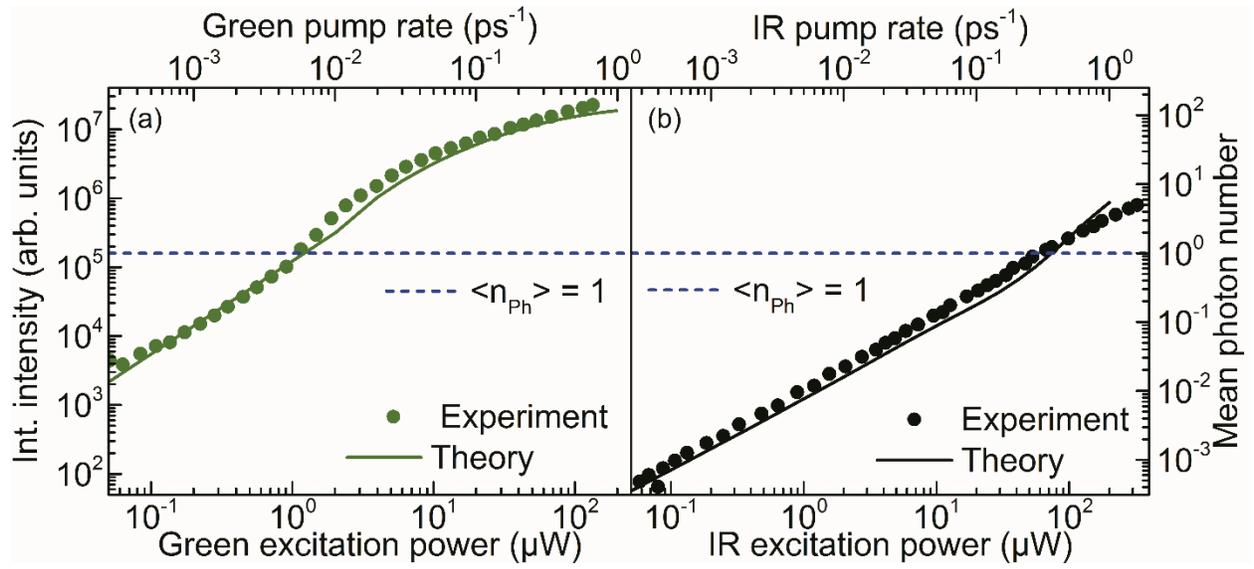

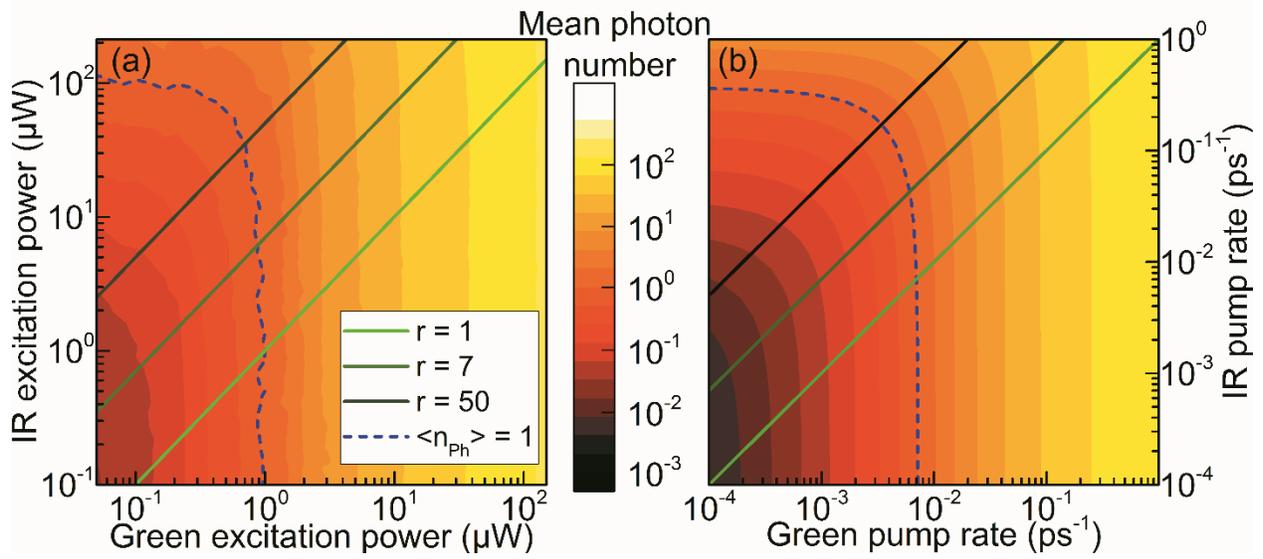

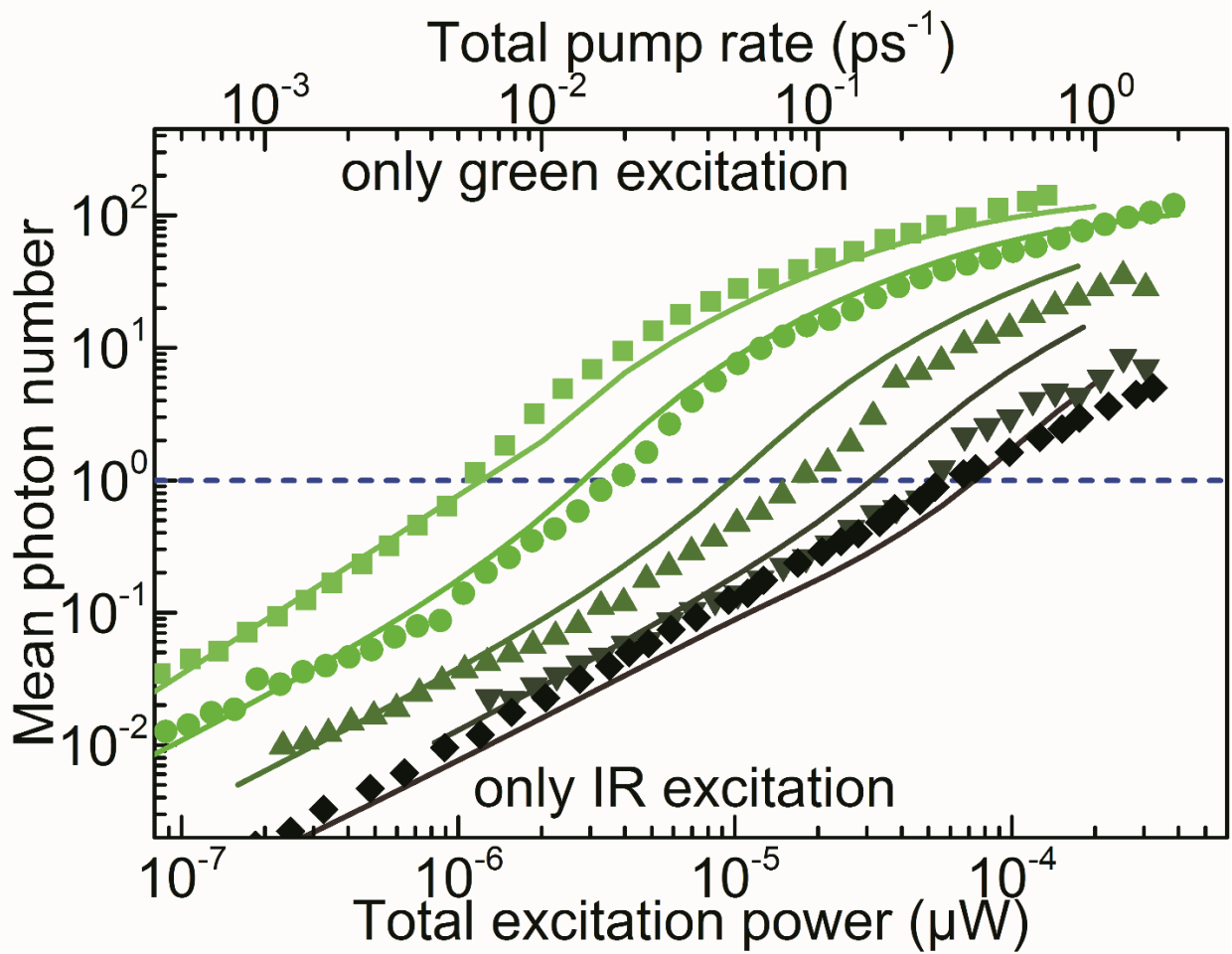

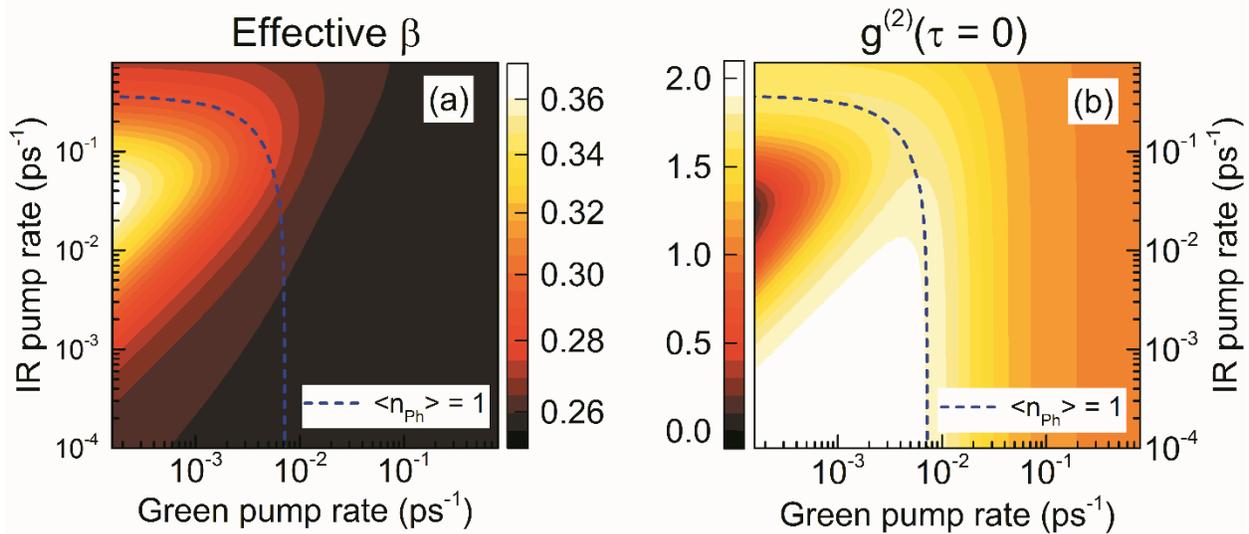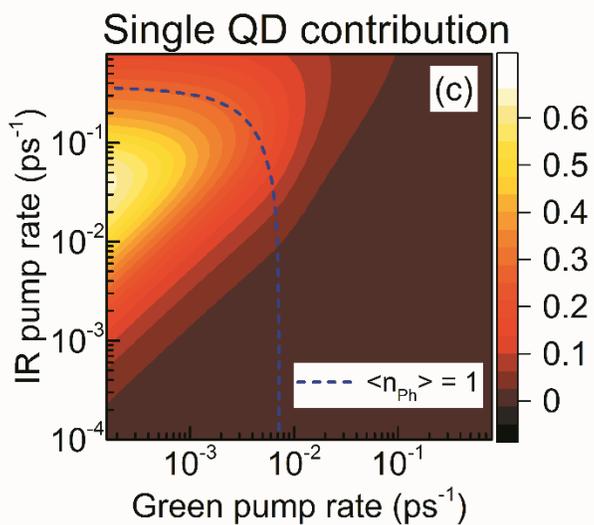